\documentclass[fp]{jpsj2}
\usepackage{txfonts}
\usepackage{bm}

\title{Correlated Electrons and Mass Enhancement on the $\beta$-Mn Type Lattice}

\author{Yoshiro Kakehashi\thanks{yok@sci.u-ryukyu.ac.jp, to be published in J. Phys. Soc. Jpn.}}
\inst{Department of Physics, Faculty of Science,\\
University of the Ryukyus, \\
1 Senbaru, Nishihara, Okinawa, 903-0213, Japan} %\\

\abst{\vspace*{3mm}
Electronic structure and electron correlations of the Hubbard model on the $\beta$-Mn lattice have been investigated on the basis of the momentum dependent local ansatz wavefunction method.  It is found that the electrons on the $\beta$-Mn lattice are significantly correlated as compared with those on the face centered cubic (fcc) and the body centered cubic (bcc) lattices; the correlation energy gain of the $\beta$-Mn lattice structure is the largest among the $\beta$-Mn, the fcc, and the bcc structures, and the charge fluctuations show a significant suppression.  In particular, it is found that the mass enhancement factor on the $\beta$-Mn lattice shows a large peak around the electron number per atom $n=1.1$ when $n$ is varied at the typical Coulomb interaction energy strength of transition metals.  It is pointed that the large mass enhancement may explain anomalous electronic specific heat coefficient in $\beta$-Mn.
}

\kword{ beta-Mn, mass enhancement, electronic specific heat, variational method, momentum-dependent local ansatz, electron correlations, momentum distribution function, transition metal }

\begin{document}
\maketitle

\section{Introduction}

Physical property of transition metals and alloys has been much investigated over the past century.~\cite{guill97,bozor51,slater36,vanv53}  Most of the electronic and magnetic properties at the ground state have been quantitatively explained on the basis of the density functional theory for band calculations.~\cite{hohen64,kohn65,moruzzi78,moruzzi95}  Nevertheless some of 3d transition metals and compounds have not yet been well understood.
$\beta$-Mn is one of such metals.  There is no convincing theory to explain the stability of the paramagnetism~\cite{kasper56,masuda64} and a large electronic specific heat of this system~\cite{shinkoda79} on the basis of the electronic structure.

$\beta$-Mn shows the cubic A13 structure with 20 atoms in a unit cell~\cite{preston28} whose space group is given by P4${}_{1}$32.  There are two inequivalent sites, site I (the 8c site) and site II (the 12d site).  Among 20 atoms, 8 atoms belong to site I, and the remaining 12 atoms to site II.
Early band calculations based on the local spin density approximation by Sliwko {\it et. al.}~\cite{sliwko94} reported the nonmagnetic ground state at the experimental volume. Asada~\cite{asada95} performed the LMTO (linear muffin-tin orbital) calculations with use of the generalized gradient approximation potential, and obtained a complex antiferromagnetic structure in which the Mn I moments vanish and the magnitudes of Mn II moments are 2.0 $\mu_{\rm B}$ at the experimental volume.  Later, Hafner and Hobbs~\cite{hafner03} obtained the ferrimagnetic structure with the Mn I moments $-0.8$ $\mu_{\rm B}$ and the Mn II moments $1.6$ $\mu_{\rm B}$ at the same volume.
It was also found that the calculated density of states (DOS) at the Fermi level is one order of magnitude lower than that deduced from the specific heat measurement, indicating that the observed mass enhancement factor of $\beta$-Mn $m^{\ast}/m \sim 10$.

The enhancement of the electronic specific heat coefficient $\gamma$ in $\beta$-Mn is usually explained by the nearly antiferromagnetic spin fluctuations~\cite{shinkoda79,hase75} based on the self-consistent spin fluctuation theory~\cite{moriya73,moriya85}.  Although the $\gamma$ vs N\'{e}el temperature curves of $\beta$-Mn alloys seem to be consistent with the experimental data~\cite{shinkoda79}, the electronic reason for the paramagnetism in $\beta$-Mn and the role of the $\beta$-Mn type lattice structure have not yet been clarified.

To explain the paramagnetism of $\beta$-Mn, Nakamura {\it et. al.}~\cite{naka97} proposed the suppression of the magnetic order due to strong antiferromagnetic spin frustrations of Mn II local moments on a network of corner sharing triangles being characteristic of the $\beta$-Mn lattice.
The inelastic neutron scattering experiments showing strong antiferromagnetic spin fluctuations seem to be consistent with the spin frustration picture~\cite{shiga95,paddi13}.
The frustration picture of Mn II local moments, however, is not consistent with the electronic state~\cite{hafner03} with large nearest-neighbor (NN) electron hopping integrals between site I and site II.

In order to understand the physics of $\beta$-Mn from the microscopic point of view, it is desirable to examine the physical properties on the basis of the model which takes into account the electronic structure on the $\beta$-Mn type lattice.
In this paper, we consider a simple Hubbard model~\cite{gutz63,hubbard63} on the $\beta$-Mn lattice with the transfer integrals up to the 6 NN, and clarify the electronic structure and electron correlations at the paramagnetic ground state on the basis of the momentum dependent local ansatz (MLA) wavefunction method.~\cite{kake08,pat11,pat13,pat13-2,kake14}
The MLA takes into account the two particle excited states expanded by the residual Coulomb interactions, and describes best the correlated electrons in the metallic regime, introducing momentum dependent variational parameters.

We will demonstrate that the electrons on the $\beta$-Mn type lattice are significantly correlated as compared with those on the face centered cubic (fcc) lattice and the body centered cubic (bcc) lattice.  In particular we found a large mass enhancement factor when the electron number is increased by 0.1 from the half filling, because of the band narrowing and the existence of a characteristic peak in the DOS.  The result indicates that a large electronic specific heat in $\beta$-Mn may be explained by the Mott-Hubbard type mass enhancement.

In the following section, we outline the MLA for the system with more than one atom in a unit cell.
In Sect. 3, we present numerical results for the $\beta$-Mn lattice.  Calculated DOS as well as the local densities of states (LDOS) on the $\beta$-Mn lattice are shown to be narrow as compared with those of the fcc and the bcc.  The correlation energy, the charge fluctuations, the momentum distribution function, and the mass enhancement factor are obtained as a function of the electron number per atom $n$ and the Coulomb interaction energy parameter $U$.  A large peak of mass enhancement factor is obtained at the electron number $n=1.1$ for the interaction $U$ being typical to the 3d transition metals.
The results are summarized in the last section.  It is suggested that the strong mass enhancement obtained in the present calculations may explain the experimental specific heat anomaly in $\beta$-Mn.  Discussions on the quantitative calculations and the stability of the paramagnetism are given.

\section{Momentum Dependent Local Ansatz Wavefunction Method}

We consider the single-band Hubbard model~\cite{gutz63,hubbard63} with more than one atom in a unit cell.
\begin{equation}
H =  
\sum_{i\sigma}\epsilon_{i} \, n_{i\sigma} + \sum_{ij\sigma} t_{ij} \, a_{i\sigma}^{\dagger} a_{j\sigma}     
+ \sum_{i} U_{i} \, n_{i\uparrow} n_{i\downarrow} \ .
\label{hmodel}
\end{equation}
Here $\epsilon_{i}$ is the atomic level on site $i$, 
$t_{ij}$ is the transfer integral between sites $i$ and $j$, and $U_{i}$
is the intra-atomic Coulomb interaction energy parameter on site $i$.  
$a_{i \sigma}^{\dagger}$ ($a_{i \sigma}$) denotes the creation (annihilation) operator for an
electron on site $i$ with spin $\sigma$, and 
$n_{i\sigma}=a_{i\sigma}^{\dagger} a_{i\sigma}$ is the electron density
operator on site $i$ for spin $\sigma$.

In the Hartree-Fock approximation, we neglect the fluctuation terms for the Coulomb interactions and replace the original Hamiltonian (\ref{hmodel}) with an independent electron  Hamiltonian $H_{0}$ as follows.
\begin{equation}
H_{0}= \sum_{i\sigma} \epsilon_{i\sigma} \, 
n_{i\sigma} + \sum_{ij\sigma} t_{ij} \, a_{i\sigma}^\dagger a_{j\sigma}
- \sum_{i} U_{i} \, \langle n_{i\uparrow}\rangle_0\langle 
n_{i\downarrow}\rangle_0 \ . 
\label{hfh0}
\end{equation}
Here $\epsilon_{i\sigma}=\epsilon_{i}+U_{i} \, \langle n_{i-\sigma}\rangle_0$, and 
$\langle \sim \rangle_0$ denotes the average with respect to the Hartree-Fock ground state $|\phi_{0} \rangle$.

The original Hamiltonian (\ref{hmodel}) is expressed with use of $H_{0}$ as follows.
\begin{equation}
H = H_{0} + \sum_{i} U_{i} O_{i} \ .
\label{hmodel2}
\end{equation}
Here $O_{i}=\delta n_{i\uparrow} \delta n_{i\downarrow}$ is the residual Coulomb interaction and $\delta n_{i\sigma}$ is defined
by $\delta n_{i\sigma}=n_{i\sigma}-\langle n_{i\sigma} \rangle_0$ \ .

The Hartree-Fock Hamiltonian (\ref{hfh0}) is diagonalized as follows. 
\begin{equation}
H_{0}= \sum_{\kappa\sigma} \epsilon_{\kappa\sigma} \, 
n_{\kappa\sigma} - \sum_{i} U_{i} \, \langle n_{i\uparrow}\rangle_0\langle 
n_{i\downarrow}\rangle_0 \ . 
\label{hfdiag}
\end{equation}
Here $\epsilon_{\kappa\sigma}$ are the eigen values for the one-electron Hamiltonian 
$H_{ij\sigma}=\epsilon_{i\sigma}\delta_{ij}+t_{ij}$ and $\kappa$ denotes the momentum $\boldsymbol{k}$ 
and the branch number $\nu$, {\it i.e.}, $\kappa=\boldsymbol{k}\nu$.  We assume hereafter that the 
energy $\epsilon_{\kappa\sigma}$ is measured from the Fermi level $\epsilon_{F}$,  
introducing the chemical potential term $-\epsilon_{F} N$ into Eq. (\ref{hfh0}).  
Here $N$ is the number of electrons.
Moreover $n_{\kappa\sigma}=a_{\kappa\sigma}^{\dagger}a_{\kappa\sigma}$ is the electron 
density operator for an electron with the eigen state $\kappa$ and spin $\sigma$, 
and $a_{\kappa\sigma}^{\dagger}$ ($a_{\kappa\sigma}$) denotes the creation (annihilation) 
operator for the same electron.

The Hartree-Fock ground state energy is given by
\begin{equation}
\langle H \rangle_{0} = \langle H_{0} \rangle_{0} = \sum_{\kappa\sigma}^{\rm occ} \epsilon_{\kappa\sigma} 
 - \sum_{i} U_{i} \, \langle n_{i\uparrow} \rangle_{0} \langle 
n_{i\downarrow} \rangle_{0} \ .
\label{hfgener}
\end{equation}
Note that the sum over $\kappa\sigma$ is taken with respect to the occupied (occ) electrons.  The Hartree-Fock ground state is expressed as
\begin{equation}
|\phi_0\rangle=\left[\prod^{\rm occ}_{\kappa}  a^{\dagger}_{\kappa\uparrow} \right]
\left[\prod^{\rm occ}_{\kappa}  a^{\dagger}_{\kappa\downarrow} \right] |0 \rangle \ .
\label{hfphi}
\end{equation}
Here $|0\rangle$ denotes the vacuum state.

The Hartree-Fock approximation does not describe electron correlations.
In order to take into account the correlations at zero temperature, we adopt the momentum dependent local ansatz (MLA) wavefunction~\cite{pat11}.
\begin{eqnarray}
|\Psi_{\rm MLA} \rangle = \left[ \prod_{i} (1 - \tilde{O}_{i}) \right] |\phi_{0} \rangle \ .
\label{mlawf}
\end{eqnarray}
Here $\tilde{O}_{i}$ is the local correlator on site $i$ with momentum-dependent variational parameters $\eta^{(i)}_{\kappa^{\prime}_{2} \kappa_{2} \kappa^{\prime}_{1} \kappa_{1}}$, which is given by
\begin{eqnarray}
\tilde{O}_{i} = \sum_{\kappa_{1} \kappa_{2} \kappa^{\prime}_{1} \kappa^{\prime}_{2}} 
\langle \kappa^{\prime}_{2} | i \rangle_{\downarrow} \langle i | \kappa_{2} \rangle_{\downarrow} 
\langle \kappa^{\prime}_{1} | i \rangle_{\uparrow} \langle i | \kappa_{1} \rangle_{\uparrow} \, 
\eta^{(i)}_{\kappa^{\prime}_{2} \kappa_{2} \kappa^{\prime}_{1} \kappa_{1}} 
\delta(a^{\dagger}_{\kappa^{\prime}_{2}\downarrow} a_{\kappa_{2}\downarrow})
\delta(a^{\dagger}_{\kappa^{\prime}_{1}\uparrow} a_{\kappa_{1}\uparrow}) \ .
\label{otilde}
\end{eqnarray}
Note that the overlap matrix $\langle i | \kappa \rangle_{\sigma}$ between the local orbital 
$i$ and the one electron eigen state $\kappa$ is a projector from $\kappa$ to $i$, so that 
$\tilde{O}_{i}$ reduces to $O_{i}=\delta n_{i\uparrow} \delta n_{i\downarrow}$ when 
$\eta^{(i)}_{\kappa^{\prime}_{2} \kappa_{2} \kappa^{\prime}_{1} \kappa_{1}}$ are momentum independent, {\it e.g.}, $\eta^{(i)}_{\kappa^{\prime}_{2} \kappa_{2} \kappa^{\prime}_{1} \kappa_{1}} = 1$.

The MLA wavefunction reproduces the result of the second-order perturbation theory in the 
weak Coulomb interaction limit and describes  well the electron correlations in the metallic regime.

The variational parameters $\eta^{(i)}_{\kappa^{\prime}_{2} \kappa_{2} \kappa^{\prime}_{1} \kappa_{1}}$ are obtained by minimizing the ground-state correlation energy $E_{c}$:
\begin{equation}
E_{c} = \langle H \rangle- \langle H \rangle_{0} =
\frac{ \langle \Psi_{\rm MLA} |\tilde{H}| \Psi_{\rm MLA} \rangle }
{ \langle \Psi_{\rm MLA} | \Psi_{\rm MLA} \rangle } \ . 
\label{corr1}
\end{equation}
Here \ \ $\tilde{H} = H - \langle H \rangle_{0}$.
In the single-site approximation (SSA), the correlation energy is given by
\begin{equation}
E_{c} = \sum_{i} \epsilon^{(i)}_{c}  \ .
\label{corr2}
\end{equation}
The correlation energy $\epsilon^{(i)}_{c}$ on each site $i$ is given by
\begin{equation}
\epsilon^{(i)}_{c} =\frac{  - \langle \tilde{O}^{\dagger}_{i} \tilde{H} \rangle_0 
- \langle \tilde{H}  \tilde{O}_{i} \rangle_0 
+ \langle \tilde{O}^{\dagger}_{i} \tilde{H} \tilde{O}_{i} \rangle_0 } 
{1+ \langle \tilde{O}^{\dagger}_{i}  \tilde{O}_{i} \rangle_0 } . 
\label{corr3}
\end{equation}
The matrix elements at the rhs (right-hand-side) of Eq. (\ref{corr3}) can be calculated with use of Wick's theorem as given in Appendix.

The self-consistent equations for variational parameters $\eta^{(i)}_{\kappa^{\prime}_{2} \kappa_{2} \kappa^{\prime}_{1} \kappa_{1}}$ are obtained from the variational principle 
$\delta E_{c} = \sum_{i} \delta \epsilon^{(i)}_{c} = 0$.  Since it is not easy to solve the equations, 
we make use of a simplified ansatz for the variational parameters which is exact in 
the weak Coulomb interaction limit and in the atomic limit.
\begin{eqnarray}
\eta^{(i)}_{\kappa^{\prime}_{2} \kappa_{2} \kappa^{\prime}_{1} \kappa_{1}} = 
\dfrac{U_{i} \, \tilde{\eta}_{i}}
{\Delta E_{{\kappa^{\prime}_{2}\kappa_{2}\kappa^{\prime}_{1}\kappa_{1}}} - \epsilon^{(i)}_{\rm c}} \ .
\label{eta2}
\end{eqnarray}
Here 
$\Delta E_{\kappa^{\prime}_{2} \kappa_{2} \kappa^{\prime}_{1} \kappa_{1}} 
= \epsilon_{\kappa^{\prime}_{2} \downarrow} - \epsilon_{\kappa_{2} \downarrow}
+ \epsilon_{\kappa^{\prime}_{1} \uparrow} - \epsilon_{\kappa_{1} \uparrow}$ and 
$\tilde{\eta}_{i}$ is a renormalization factor as a new variational parameter.

Substituting Eq. (\ref{eta2}) into each term at the rhs of Eq. (\ref{corr3}), we obtain
\begin{align} 
\langle \tilde{H} \tilde{O}_{i} \rangle_{0} =
\langle \tilde{O}^{\dagger}_{i} \tilde{H} \rangle_{0}^{\ast} = A_{i} U^{2}_{i} \, \tilde{\eta}_{i}
\label{ho2}
\end{align}
\begin{align} 
\langle \tilde{O}^{\dagger}_{i} \tilde{H} \tilde{O}_{i} \rangle_{0} = B_{i} U^{2}_{i} \, \tilde{\eta}^{2}_{i}
\label{oho2}
\end{align}
\begin{align} 
\langle \tilde{O}^{\dagger}_{i} \tilde{O}_{i} \rangle_{0} = C_{i} U^{2}_{i} \, \tilde{\eta}^{2}_{i}
\label{oo2}
\end{align}

The coefficients $A_{i}$, $B_{i} = B^{(1)}_{i} + U_{i} B^{(2)}_{i}$, and $C_{i}$ are given as follows.
\begin{align} 
A_{i} =
\int \frac{\Big[ \prod \limits^4_{n=1} d\epsilon_{n} \Big] \,
\rho_{i\uparrow}(\epsilon_{1}) \rho_{i\uparrow}(\epsilon_{2})
\rho_{i\downarrow}(\epsilon_{3}) \rho_{i\downarrow}(\epsilon_{4})
\tilde{f}(\epsilon_{1}, \epsilon_{2}, \epsilon_{3}, \epsilon_{4}) }
{ \epsilon_{4} - \epsilon_{3} + \epsilon_{2} - \epsilon_{1} -\epsilon^{(i)}_{c} } \ ,
\label{ohai}
\end{align}
\begin{align}
B^{(1)}_{i} =  
\int \frac{\Big[ \prod \limits^4_{n=1} d\epsilon_{n}\Big] \, 
\rho_{i\uparrow}(\epsilon_{1}) \rho_{i\uparrow}(\epsilon_{2})
\rho_{i\downarrow}(\epsilon_{3}) \rho_{i\downarrow}(\epsilon_{4})
\tilde{f}(\epsilon_{1}, \epsilon_{2}, \epsilon_{3}, \epsilon_{4}) \, 
(\epsilon_{4} - \epsilon_{3} + \epsilon_{2} - \epsilon_{1}) }
{(\epsilon_{4} - \epsilon_{3} + \epsilon_{2} - \epsilon_{1} -\epsilon^{(i)}_{c})^{2} } \ ,
\label{ohob1}
\end{align}
\begin{align}
B^{(2)}_{i} & = 
\int \frac{ \Big[ \prod \limits^4_{n=1} d\epsilon_{n} \Big] \, 
\rho_{i\uparrow}(\epsilon_{1}) \rho_{i\uparrow}(\epsilon_{2})
\rho_{i\downarrow}(\epsilon_{3}) \rho_{i\downarrow}(\epsilon_{4})
\tilde{f}(\epsilon_{1}, \epsilon_{2}, \epsilon_{3}, \epsilon_{4}) }
{ \epsilon_{4} - \epsilon_{3} + \epsilon_{2} - \epsilon_{1} -\epsilon^{(i)}_{c} } \nonumber \\
& \hspace*{10mm} \times \Bigg[ \ 
\int \frac{ d\epsilon_{5} d\epsilon_{6} \, 
\rho_{i\uparrow}(\epsilon_{5}) \rho_{i\downarrow}(\epsilon_{6}) 
f(\epsilon_{5}) f(\epsilon_{6}) }
{ \epsilon_{4} - \epsilon_{6} + \epsilon_{2} - \epsilon_{5} -\epsilon^{(i)}_{c} } \nonumber \\
& \hspace*{17mm} - \int \frac{ d\epsilon_{5} d\epsilon_{6} \, 
\rho_{i\uparrow}(\epsilon_{5}) \rho_{i\downarrow}(\epsilon_{6}) 
f(-\epsilon_{5}) f(\epsilon_{6}) }
{ \epsilon_{4} - \epsilon_{6} + \epsilon_{5} - \epsilon_{1} -\epsilon^{(i)}_{c} } \nonumber \\
& \hspace*{20mm} - \int \frac{ d\epsilon_{5} d\epsilon_{6} \, 
\rho_{i\uparrow}(\epsilon_{5}) \rho_{i\downarrow}(\epsilon_{6}) 
f(\epsilon_{5}) f(-\epsilon_{6}) }
{ \epsilon_{6} - \epsilon_{3} + \epsilon_{2} - \epsilon_{5} -\epsilon^{(i)}_{c} } \nonumber \\
& \hspace*{23mm} + \int \frac{ d\epsilon_{5} d\epsilon_{6} \, 
\rho_{i\uparrow}(\epsilon_{5}) \rho_{i\downarrow}(\epsilon_{6}) 
f(-\epsilon_{5}) f(-\epsilon_{6}) }
{ \epsilon_{6} - \epsilon_{3} + \epsilon_{5} - \epsilon_{1} -\epsilon^{(i)}_{c} }  \ \ \Bigg] \ ,
\label{ooob2}
\end{align}
\begin{align} 
C_{i} =  
\int \frac{ \Big[ \prod \limits^4_{n=1} d\epsilon_{n} \Big] \,  
\rho_{i\uparrow}(\epsilon_{1}) \rho_{i\uparrow}(\epsilon_{2})
\rho_{i\downarrow}(\epsilon_{3}) \rho_{i\downarrow}(\epsilon_{4})
\tilde{f}(\epsilon_{1}, \epsilon_{2}, \epsilon_{3}, \epsilon_{4}) }
{(\epsilon_{4} - \epsilon_{3} + \epsilon_{2} - \epsilon_{1} -\epsilon^{(i)}_{c})^{2} } \ .
\label{oo2}
\end{align}
Here $\tilde{f}(\epsilon_{1}, \epsilon_{2}, \epsilon_{3}, \epsilon_{4}) = f(\epsilon_{1}) 
f(-\epsilon_{2}) f(\epsilon_{3}) f(-\epsilon_{4})$, $f(\epsilon)$ is the Fermi distribution function at zero temperature, and $\rho_{i\sigma}(\epsilon)$ is the LDOS for spin $\sigma$ on site $i$, being defined by $\rho_{i\sigma}(\epsilon)=\sum_{\kappa} | \langle i | \kappa \rangle_{\sigma} |^{2} \delta(\epsilon - \epsilon_{\kappa\sigma})$.

The ground-state energy $E$ is then given as
\begin{equation}
E = \epsilon_{F}N + \langle H \rangle_{0} + \sum_{i} \epsilon^{(i)}_{c}  \ ,
\label{gener}
\end{equation}
\begin{equation}
\epsilon^{(i)}_{c} =\frac{  - 2A_{i} U_{i}^{2} \tilde{\eta}_{i} + B_{i} U_{i}^{2} \tilde{\eta}^{2}_{i} } 
{1+ C_{i} U_{i}^{2} \tilde{\eta}^{2}_{i} } \ . 
\label{corr4}
\end{equation}
The variational parameter $\tilde{\eta}_{i}$ is easily obtained from the variational principle 
$\delta E = \sum_{i} \delta \epsilon^{(i)}_{c} =0$ as
\begin{equation}
\tilde{\eta}_{i} = \frac{  - B_{i} + \sqrt{ B^{2}_{i} + 4 A_{i}^{2} C_{i} U^{2}_{i} } } 
{ 2A_{i}C_{i} U_{i}^{2} } \ . 
\label{eta3}
\end{equation}

The electron number $\langle n_{i} \rangle$ on site $i$ is given as follows.
\begin{equation}
\langle n_{i} \rangle = \langle n_{i} \rangle_0 \, + \,
\frac{\displaystyle \ \sum_{\sigma} D_{i\sigma} U^{2}_{i} \, \tilde{\eta}^{2}_{i} \ }
{\displaystyle  1 + C_{i} U_{i}^{2} \tilde{\eta}^{2}_{i} } \ . 
\label{aveni}
\end{equation}
Here $\langle n_i \rangle_0$ is the electron number in the Hartree-Fock approximation, 
and
\begin{align} 
D_{i\sigma} &= 
\int \frac{ \Big[ \prod \limits^4_{n=1} d\epsilon_{n} \Big] \, 
\rho_{i\sigma}(\epsilon_{1}) \rho_{i\sigma}(\epsilon_{2})
\rho_{i -\sigma}(\epsilon_{3}) \rho_{i -\sigma}(\epsilon_{4})
\tilde{f}(\epsilon_{1}, \epsilon_{2}, \epsilon_{3}, \epsilon_{4}) }
{ \epsilon_{4} - \epsilon_{3} + \epsilon_{2} - \epsilon_{1} -\epsilon^{(i)}_{c} } 
\nonumber \\
&\hspace{10mm} \times \Bigg[ 
\int \frac{ d\epsilon_{5} \, \rho_{i\sigma}(\epsilon_{5}) f(-\epsilon_{5}) }
{ \epsilon_{4} - \epsilon_{3} + \epsilon_{5} - \epsilon_{1} -\epsilon^{(i)}_{c} }
- \int \frac{ d\epsilon_{5} \, \rho_{i\sigma}(\epsilon_{5}) f(\epsilon_{5}) }
{ \epsilon_{4} - \epsilon_{3} + \epsilon_{2} - \epsilon_{5} -\epsilon^{(i)}_{c} } \ \Bigg] \ .
\label{ono2}
\end{align}
Electron number per atom $n$ is given by
\begin{equation}
n = \frac{1}{L} \sum_{i} \langle n_{i} \rangle  \ . 
\label{aven}
\end{equation}
Here $L$ is the number of sites.

Equations (\ref{corr4}), (\ref{eta3}),  (\ref{aveni}), and (\ref{aven}) form the self-consistent equations for 
$\tilde{\eta}_{i}$, $\epsilon^{(i)}_{c}$, and $\epsilon_{F}$.
Note that the electronic structure of the system is taken into account via the LDOS $\rho_{i\sigma}(\epsilon)$ in the SSA-MLA.

Momentum distribution function (MDF) $\langle n_{\kappa\sigma} \rangle$ is given as follows.
\begin{equation}
\langle n_{\kappa\sigma} \rangle = f(\epsilon_{\kappa \sigma}) + 
\sum_{i} | \langle \kappa | i \rangle_{\sigma} |^{2} \, 
\frac{ \lambda^{(i)}_{1 -\sigma}(\epsilon_{\kappa \sigma}) f(-\epsilon_{\kappa \sigma})
- \lambda^{(i)}_{2 -\sigma}(\epsilon_{\kappa \sigma}) f(\epsilon_{\kappa \sigma})  }
{ 1 + C_{i} U_{i}^{2} \tilde{\eta}^{2}_{i} } \ .
\label{nkdis}
\end{equation}
The functions $\lambda^{(i)}_{1 -\sigma}(\epsilon_{\kappa \sigma})$ and 
$\lambda^{(i)}_{2 -\sigma}(\epsilon_{\kappa \sigma})$ denote the particle and hole excitations, respectively :
\begin{align} 
\lambda^{(i)}_{1 -\sigma}(\epsilon_{\kappa \sigma}) =  U_{i}^{2} \tilde{\eta}^{2}_{i} 
\int \frac{ d\epsilon_{1}d\epsilon_{3}d\epsilon_{4} \,  
\rho_{i\sigma}(\epsilon_{1}) \rho_{i -\sigma}(\epsilon_{3})
\rho_{i -\sigma}(\epsilon_{4}) 
 f(\epsilon_{1}) f(\epsilon_{3}) f(-\epsilon_{4}) }
{(\epsilon_{4} - \epsilon_{3} + \epsilon_{\kappa\sigma} - \epsilon_{1} -\epsilon^{(i)}_{c})^{2} } \ ,
\label{lambda1}
\end{align}
\begin{align} 
\lambda^{(i)}_{2 -\sigma}(\epsilon_{\kappa \sigma}) =  U_{i}^{2} \tilde{\eta}^{2}_{i} 
\int \frac{ d\epsilon_{2}d\epsilon_{3}d\epsilon_{4} \,  
\rho_{i\sigma}(\epsilon_{2}) \rho_{i -\sigma}(\epsilon_{3})
\rho_{i -\sigma}(\epsilon_{4}) 
 f(-\epsilon_{2}) f(\epsilon_{3}) f(-\epsilon_{4}) }
{(\epsilon_{4} - \epsilon_{3} + \epsilon_{2} - \epsilon_{\kappa\sigma}  - \epsilon^{(i)}_{c})^{2} } \ .
\label{lambda2}
\end{align}

The MDF (\ref{nkdis}) depends on the eigenstate via 
$| \langle \kappa | i \rangle_{\sigma} |^{2}$.  Taking the average of 
$| \langle \kappa | i \rangle_{\sigma} |^{2}$ over energy surface $\epsilon_{\kappa\sigma}$,
we obtain a simpler expression of $\langle n_{\kappa\sigma} \rangle$:
\begin{equation}
\langle n_{\kappa\sigma} \rangle = f(\epsilon_{\kappa \sigma}) + 
\sum_{i} \frac{ \rho_{i\sigma}(\epsilon_{\kappa\sigma}) } { \rho_{\sigma}(\epsilon_{\kappa\sigma}) } \, 
\frac{ \lambda^{(i)}_{1 -\sigma}(\epsilon_{\kappa \sigma}) f(-\epsilon_{\kappa \sigma})
- \lambda^{(i)}_{2 -\sigma}(\epsilon_{\kappa \sigma}) f(\epsilon_{\kappa \sigma})  }
{ 1 + C_{i} U_{i}^{2} \tilde{\eta}^{2}_{i} } \ .
\label{avnk}
\end{equation}
Here $\rho_{\sigma}(\epsilon) = \sum_{i} \rho_{i\sigma}(\epsilon)$ is the total density of states for the electrons with spin $\sigma$.

The average quasiparticle weight $Z_{\sigma}$ is obtained from Eq.(\ref{avnk}), taking the difference at the Fermi level.
\begin{equation}
Z_{\sigma} = 1 -  
\sum_{i} \frac{ \rho_{i\sigma}(0) } { \rho_{\sigma}(0) } \,
\frac{ \lambda^{(i)}_{1 -\sigma}(0) + \lambda^{(i)}_{2 -\sigma}(0) }
{ 1 + C_{i} U_{i}^{2} \tilde{\eta}^{2}_{i} } \ .
\label{qpz}
\end{equation}
The mass enhancement factor $\left( m^{\ast}/m \right)_{\sigma}$ for spin $\sigma$ 
is obtained from the relation $\left( m^{\ast}/m \right)_{\sigma}=1/Z_{\sigma}$, and the average 
mass enhancement factor $m^{\ast}/m$ is given by
\begin{align}
\frac{m^{\ast}}{m} = \dfrac{\rho_{\uparrow}(0)}{\rho(0)} 
\left( \frac{m^{\ast}}{m} \right)_{\uparrow} + \, \dfrac{\rho_{\downarrow}(0)}{\rho(0)} \left( \frac{m^{\ast}}{m} \right)_{\downarrow} \, .
\label{avmeff}
\end{align}
Here $\rho(\epsilon)=\rho_{\uparrow}(\epsilon)+\rho_{\downarrow}(\epsilon)$.

The other physical quantities can also be obtained from the wavefunction.  The local charge fluctuations on site $i$, for example, is obtained from the relation
\begin{align}
\langle (\delta n_{i})^{2} \rangle = 
\langle n_{i} \rangle(1 - \langle n_{i} \rangle) 
+ 2\langle n_{i\uparrow}n_{i\downarrow} \rangle \ ,
\label{flucn}
\end{align}
and the double occupation number is given by
\begin{align}
\langle n_{i\uparrow}n_{i\downarrow} \rangle = 
\langle n_{i\uparrow} \rangle_{0} \langle n_{i\downarrow} \rangle_{0} 
 + \dfrac{ -2A_{i}U_{i}\eta_{i}   
+ (B^{(2)}_{i} 
+ \sum_{\sigma} \langle n_{i -\sigma} \rangle_{0} D_{i\sigma} ) \, U^{2}_{i} \tilde{\eta}^{2}_{i}
}
{1 + C_{i}U^{2}_{i}\tilde{\eta}^{2}_{i} } \ .
\label{dblave1}
\end{align}

The multiple integrals up to the 6 folds in Eqs. (\ref{ohai}) $\sim$ (\ref{oo2}), (\ref{ono2}), 
(\ref{lambda1}), and (\ref{lambda2}) can be reduced to the integrals  up to the 2 folds by 
using the Laplace transform.~\cite{schweitz91,pat11}

\section{Numerical Results}

We considered in the numerical calculations the Hubbard model on the $\beta$-Mn type lattice in the paramagnetic state.  The $\beta$-Mn has 20 atoms in a unit cell with two inequivalent sites, site I and site II.  
We assumed the same atomic levels for both sites I and II: $\epsilon_{i} = 0$, and took into account the transfer integrals $t_{n}$ up to the 6-th NN: $t_{1}=-1.0$ and $t_{n} = -0.6$ ($n = 2 \sim 6$).  Here and hereafter we measure the energy in unit of the NN transfer integral $|t_{1}|=1$.  The choice of the parameters $t_{n}$ in the present model is based on Heine's law~\cite{heine67,poul76,ander85} on the transfer integrals for $d$ orbitals, $t_{n} = t_{ij} \varpropto R^{-5}$.  Here $R$ is the inter-atomic distance between sites $i$ and $j$.  The transfer integrals calculated with use of the $R^{-5}$ law are given in Table I together with those of the present model.  Atomic-position data were taken from Ref. 33.

We also made the calculations for the model with the NN transfer integral ($t_{1}=-1.0$) on the fcc lattice, and that on the bcc lattice, in order to clarify the characteristics of electrons on the $\beta$-Mn lattice.
The intra-atomic Coulomb interaction energy $U$ in 3d transition metals are considered to be $U/W \sim 0.5$ for the d band width $W$.~\cite{kake17}  For the fcc and the bcc, $W = 16$ in the present model, thus we can expect that $U \sim 8$.  In the following numerical calculations we adopt  $U=7$ when we change electron number per atom $n$, bearing in mind 3d transition metal system.
%
%
%---------------------------------------------------------------------
\begin{table}[tbh]
\caption{Inter-atomic distances ($R_{n}$) and the $n$-th nearest neighbor transfer integrals ($t_{n}$).  $R_{n}$ ($|t_{n}|$) is given in unit of the lattice parameter (the nearest neighbor transfer integral $|t_{1}|$).  $|t_{n}|$ ($\propto R^{-5}_{n}$) are calculated with use of Heine's law for transition metals.~\cite{heine67,poul76,ander85}  $|t_{n}|$ (present) are used in the present model calculations.
\vspace{5mm} }
\label{table-tn}
\begin{tabular}{cccccccc}
\hline
$n$  & 1 & 2 & 3 & 4 & 5 & 6 & 7  \\ \hline
\vspace*{1mm}
$R_{n}$ & 0.3743 & 0.4080 & 0.4171 & 0.4190 & 0.4233 & 0.4243 & 0.5180 \\
$|t_{n}|$ ($\propto R^{-5}_{n}$) & 1.0 & 0.6498 & 0.5820 & 0.5689 & 0.5406 & 0.5342 & 0.1970  \\
$|t_{n}|$ (present) & 1.0 & 0.6 & 0.6 & 0.6 & 0.6 & 0.6 & 0 \\
\hline 
\end{tabular}
\end{table}
%---------------------------------------------------------------------
%
%

In the self-consistent calculations, the Fermi level $\epsilon_{F}$ is determined by Eq. (\ref{aven}), one of the self-consistent equations for the MLA.  Thus, the Hartree-Fock self-consistent equations have to be solved every time in the MLA self-consistent process.  In order to simplify the calculations, we made a frozen potential approximation in which the atomic potentials $\epsilon_{i\sigma}$ in the MLA are fixed to be those of the original Hartree-Fock potentials $\epsilon_{i\sigma}^{(hf)}$.  Then the MLA potentials $\epsilon_{i\sigma}$ are obtained only by changing the Fermi level as $\epsilon_{i\sigma} = \epsilon_{i\sigma}^{(hf)} - \Delta \mu$, where $\Delta \mu = \epsilon_{F} - \epsilon_{F}^{(hf)}$ and $\epsilon_{F}^{(hf)}$ is the Fermi level in the Hartree-Fock approximation. 
Then the input LDOS $\rho_{i\sigma}(\epsilon)$ in the self-consistent MLA equations are obtained from the LDOS $\rho_{i\sigma}^{(hf)}(\epsilon)$ in the original Hartree-Fock approximation as
\begin{eqnarray}
\rho_{i\sigma}(\epsilon) = \rho_{i\sigma}^{(hf)}(\epsilon + \Delta \mu) \ .
\label{rhoin}
\end{eqnarray}
%
%

%
%
%---------------------------------------------------------------------
\begin{figure}[htb]
\begin{center}
\includegraphics[width=10cm]{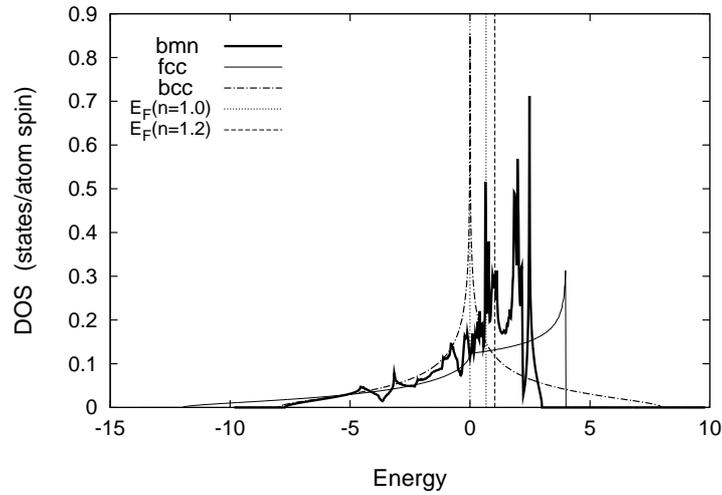}
\end{center}
%\vspace{2cm}
\caption{
Non-interacting densities of states (DOS) per atom for the $\beta$-Mn type structure (bmn: solid curve), the face centered cubic structure (fcc: thin solid curve), and the body centered cubic structure (bcc: dot-dashed curve).  The Fermi level of the bmn for $n=1.0$ ($1.2$) is also shown by a vertical dotted (dashed) line.  Here $n$ denotes the electron number per atom.
}
\label{fig-dos0}
\end{figure}
%---------------------------------------------------------------------
%
%
We present in Fig. \ref{fig-dos0} the noninteracting DOS for the fcc, the bcc, and the $\beta$-Mn (bmn) lattice structures.
The fcc DOS monotonically increases with increasing energy and shows a peak at the top of the DOS.  
The bcc DOS is symmetric with respect to the energy $\epsilon=0$ axis because of the particle-hole symmetry of the energy bands, and shows a huge peak at the center due to the flat energy band there.

The band widths of the fcc and bcc DOS are both 16.  The width of the DOS for the $\beta$-Mn structure, on the other hand, is 10.7; the energy bands of the $\beta$-Mn structure are narrower than the fcc and the bcc by about 30\%, because of less NN atoms and reduced magnitudes of the transfer integrals.
Moreover the DOS for the $\beta$-Mn lattice has more complex structure because of a large number of atoms in the unit cell.  The highest peak is located near the top of the bands as in the fcc DOS.  The second highest peak is located at energy $\epsilon \approx 2.0$.  The third peak appears around $\epsilon=1.0$.  Note that the Fermi level crosses over the third peak with increasing electron number $n$ from $1.0$ to $1.2$.
%
%
%---------------------------------------------------------------------
\begin{figure}[htb]
\begin{center}
\includegraphics[width=10cm]{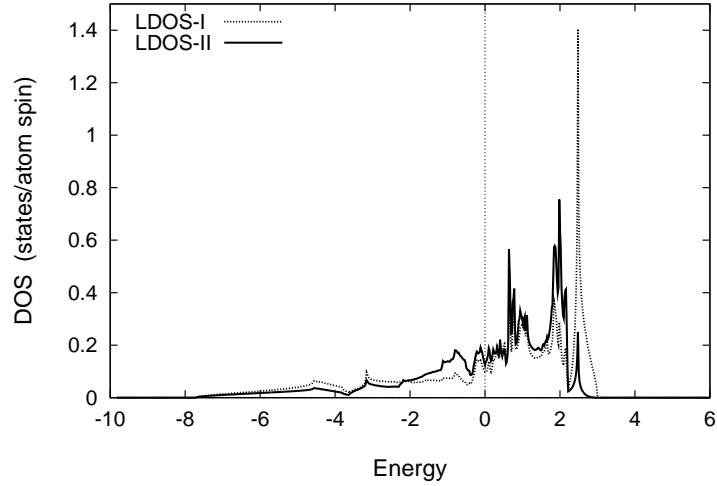}
\end{center}
%\vspace{2cm}
\caption{
The local densities of states (LDOS) for non-interacting electrons on the bmn ($\beta$-Mn) lattice.  Dotted curve (solid curve) is the LDOS on site I (site II).
}
\label{fig-ldos}
\end{figure}
%---------------------------------------------------------------------
%
%

The LDOS for site I and site II on the $\beta$-Mn type lattice are presented in Fig. \ref{fig-ldos}.  The band width of LDOS on site I is larger than that on site II because of the existence of the NN electron hoppings among site I.  It implies that the electrons on site II are more localized than on site I in general.  The second point is that the highest peak near the top of the DOS belongs to the site-I electrons, while the 2nd peak near $\epsilon=2.0$ mainly belongs to the site-II electrons.  This indicates that the spin polarization in nearly filled bands  would be caused by the electrons on site I, though the electrons on site I are more delocalized than those on site II.  It should be noted that the same properties of the DOS and LDOS on the $\beta$-Mn lattice hold true for the Hartree-Fock electrons.

Using the Hartree-Fock LDOS and Eq. (\ref{rhoin}), we solved the self-consistent equations 
(\ref{corr4}), (\ref{eta3}),  (\ref{aveni}), and (\ref{aven}) for the $\beta$-Mn lattice as well as the bcc and the fcc ones in the paramagnetic state.  Figure 3 shows the result of the self-consistent correlation energies for each structure.
The correlation energy gains $|\epsilon_{c}|$ are caused by the reduction of the double occupation number and associated reduction of electron hopping rate.  The former reduces the on-site Coulomb repulsion energy and the latter suppresses the kinetic energy gain.
We find that the correlation energy of the $\beta$-Mn lattice shows the minimum at $n=1.1$, while the bcc shows the minimum at $n=1.0$ due to the particle-hole symmetry of the system.  Furthermore the energy gain $|\epsilon_{c}|$ for the $\beta$-Mn lattice is the largest among three structures; electrons on the $\beta$-Mn lattice are more correlated than those on the other two lattice systems.
%
%
%---------------------------------------------------------------------
\begin{figure}[htb]
\begin{center}
\includegraphics[width=10cm]{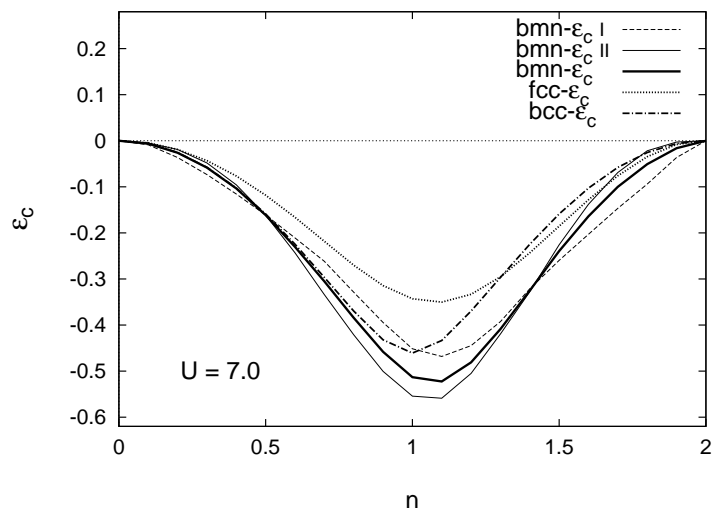}
\end{center}
%\vspace{2cm}
\caption{
Self-consistent correlation energies per atom $\epsilon_{c}$ for the $\beta$-Mn structure (bmn-$\epsilon_{c}$: solid curve), the fcc structure (fcc-$\epsilon_{c}$: dotted curve), and the bcc structure (bcc-$\epsilon_{c}$: dot-dashed curve) as a function of the electron number per atom, $n$.  The correlation energies $\epsilon^{(i)}_{c}$ of the $\beta$-Mn on sites I and II are also shown by the thin dashed curve and the thin solid curve, respectively.  The Coulomb interaction energy is fixed to be $U=7.0$.
}
\label{fig-ecn}
\end{figure}
%---------------------------------------------------------------------
%
%

\begin{figure}[htb]
\begin{center}
\includegraphics[width=10cm]{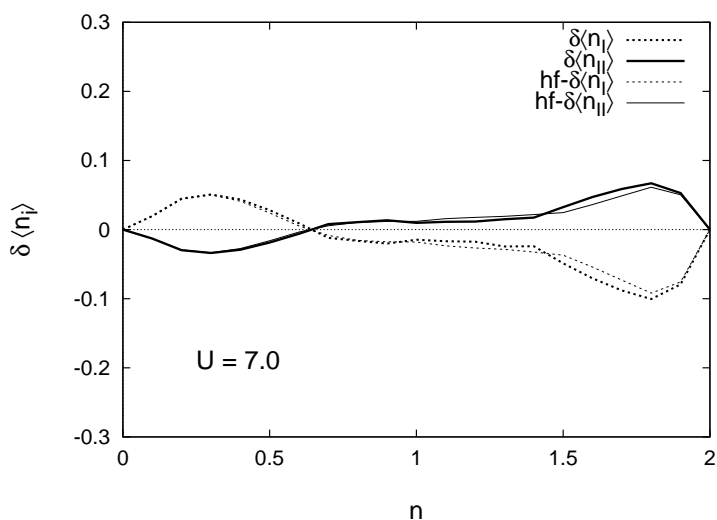}
\end{center}
%\vspace{2cm}
\caption{
Deviations from the charge neutrality on site I (dotted curve) and II (solid curve), $\delta \langle n_{i} \rangle = \langle n_{i} \rangle - n$, as a function of the electron number per atom $n$.  The Hartree-Fock curves are shown by the thin dotted curve and the thin solid curve, respectively.  The Coulomb interaction energy is fixed to be $U=7.0$.
}
\label{fig-dnn}
\end{figure}

The difference in localization of electrons between site I and site II leads to the site-dependent local properties.  Figure \ref{fig-dnn} shows the local charge polarizations measured from the neutral charge ($\delta \langle n_{i} \rangle = \langle n_{i} \rangle - n$), which are calculated in the Hartree-Fock approximation and the MLA.  Since the effective band width of the LDOS on site I is broader than that on site II, the local charge on site I is larger (smaller) than that on site II for $n < 0.6$ ($n > 0.6$); the site I is ionized negatively (positively) and the site II is ionized positively (negatively) when $n < 0.6$ ($n > 0.6$).  Correlation corrections are not significant for the charge polarization.

%
%
%---------------------------------------------------------------------
%---------------------------------------------------------------------
%
%

\begin{figure}[htb]
\begin{center}
\includegraphics[width=10cm]{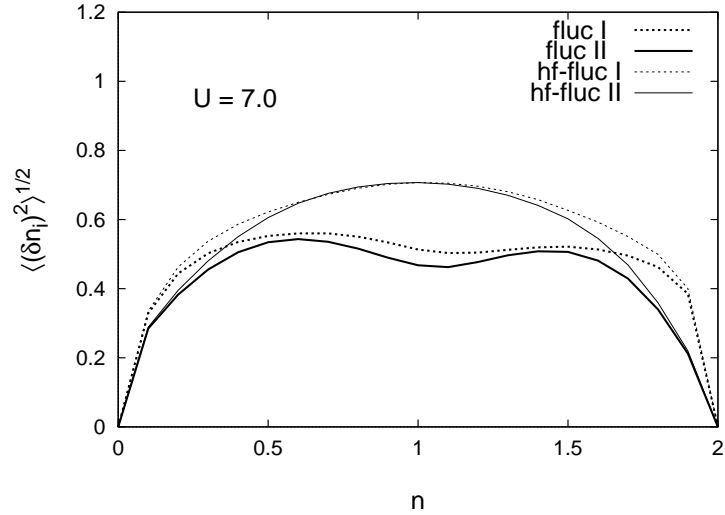}
\end{center}
%\vspace{2cm}
\caption{
Charge fluctuations $\langle \delta n_{i})^{2} \rangle^{1/2}$ on site I (fluc I: dotted curve) and site II (fluc II: solid curve) as a function of electron number per atom $n$.  The Hartree-Fock curves (hf-fluc I and hf-fluc II) are shown by thin dotted line and thin solid line, respectively.  The Coulomb interaction energy is fixed to be $U=7.0$. 
}
\label{fig-flucn}
\end{figure}

We present in Fig. \ref{fig-flucn} the charge fluctuations $\langle (\delta n_{i})^{2} \rangle^{1/2}$ as a function of electron number $n$ at $U=7.0$.  The charge fluctuations for uncorrelated electrons show a parabolic behavior according to the simple formula $\langle (\delta n_{i})^{2} \rangle = \langle n_{i} \rangle_{0} - \langle n_{i} \rangle^{2}_{0}/2$.
The larger fluctuations on site I in the region $n \lesssim 0.5$ ($n \gtrsim 1.5$) originate in more charge (hole) on site I than on site II (see Fig. \ref{fig-dnn}).

Electron correlations suppress the double occupation number and thus the charge fluctuations.  The effects are stronger on site II as seen in Fig. \ref{fig-flucn}.  
The suppression becomes maximum around $n=1.1$, and causes a dip in the curves.
%
%
%---------------------------------------------------------------------
%---------------------------------------------------------------------
%
%

\begin{figure}[htb]
\begin{center}
\includegraphics[width=10cm]{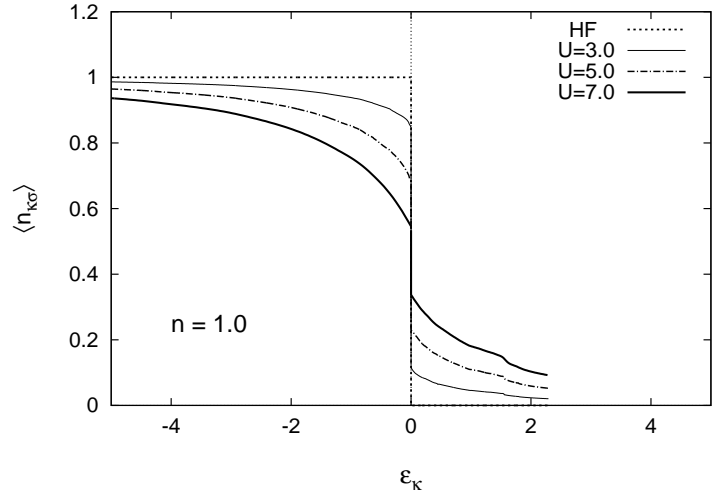}
\end{center}
%\vspace{2cm}
\caption{
Momentum distribution function $\langle n_{\kappa\sigma} \rangle$ vs energy $\epsilon_{\kappa}$ curves at half-filling for various Coulomb interaction energies $U=3.0$ (thin solid curve), $5.0$ (dot-dashed curve), and $7.0$ (solid curve).  The Fermi distribution function (, {\it i.e.}, the Hartree-Fock case) is shown by dotted curve. 
}
\label{fig-nku}
\end{figure}

We have calculated the momentum distribution function (MDF) varying the Coulomb interaction strength $U$.  Figure \ref{fig-nku} shows the average MDF curves for $U=3.0$, $5.0$, $7.0$ at half filling.  The MDF monotonically decreases with increasing energy $\epsilon_{\kappa}$ and jump at the Fermi level $\epsilon_{F} (\equiv 0)$.
The deviation from the Fermi distribution function increases with increasing $U$.
The step-like small changes in the MDF near $\epsilon_{\kappa}=1.5$ are caused by the rapid change in weight $\rho_{i\sigma}(\epsilon)/\rho_{\sigma}(\epsilon)$ at the same energy $\epsilon \approx 1.5$, which corresponds to the LDOS at the energy $\epsilon \approx 2.2$ in Fig. \ref{fig-ldos}.
%
%
%---------------------------------------------------------------------
%---------------------------------------------------------------------
%
%

\begin{figure}[htb]
\begin{center}
\includegraphics[width=10cm]{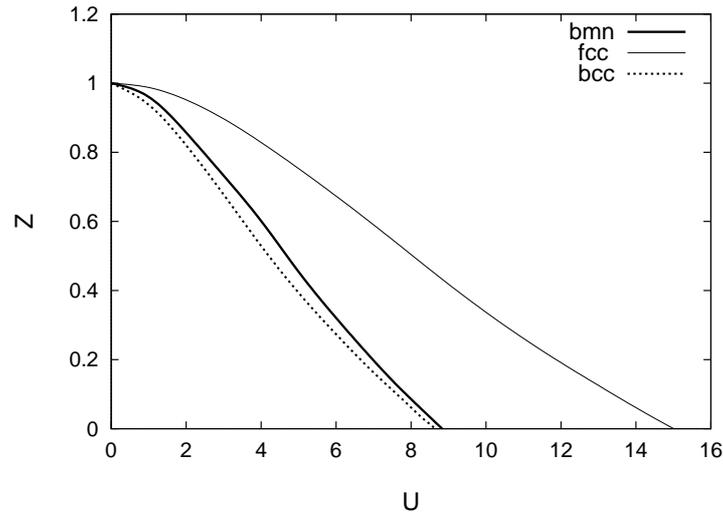}
\end{center}
%\vspace{2cm}
\caption{
Quasiparticle-weight $Z$ vs Coulomb-interaction $U$ curves at half-filling for the bmn ($\beta$-Mn) lattice (solid curve), the fcc lattice (thin solid curve), and the bcc lattice (dotted curve).
}
\label{fig-zu}
\end{figure}

The quasiparticle weight $Z$ is obtained from the jump of MDF at the Fermi level as discussed in the last section.  We present $Z$ at half-filling as a function of the Coulomb interaction energy $U$ in Fig. \ref{fig-zu}.  The quasiparticle weight $Z$ for the $\beta$-Mn lattice rapidly decreases with increasing Coulomb interaction $U$, and vanishes at the critical Coulomb interaction $U_{c2} = 8.84$, which is much smaller than $U_{c2} = 15.01$ for the fcc lattice, but is comparable to $U_{c2} = 8.63$ for the bcc.  Note that the critical value $U_{c2}$ in the MLA is generally underestimated by $10 \sim 20$\% according to the previous investigations of the metal-insulator transition in infinite dimensions.~\cite{pat11}
%
%
%---------------------------------------------------------------------
%---------------------------------------------------------------------
%
%

\begin{figure}[htb]
\begin{center}
\includegraphics[width=10cm]{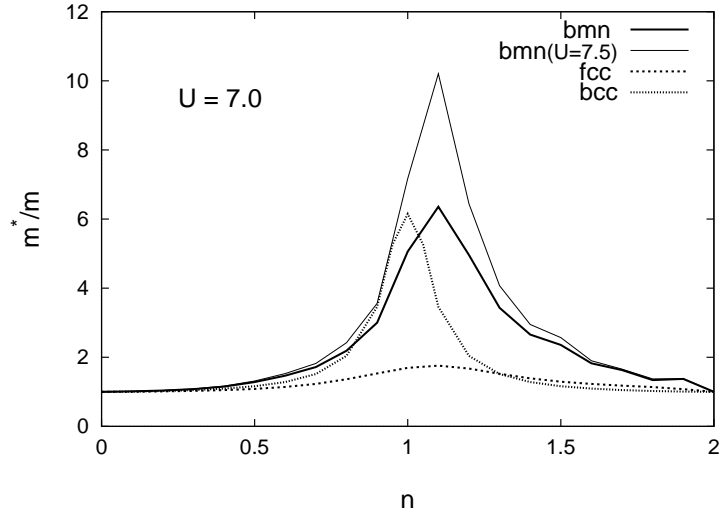}
\end{center}
%\vspace{2cm}
\caption{
Mass enhancement factors $m^{\ast}/m$ as a function of electron number $n$ at $U=7.0$ for the bmn ($\beta$-Mn) structure (solid curve), the fcc structure (dashed curve), and the bcc structure (dotted curve).  The thin solid curve shows the $m^{\ast}/m$ for the bmn at $U=7.5$.
}
\label{fig-meffn}
\end{figure}

The mass enhancement factor $m^{\ast}/m$ is obtained from the inverse quasiparticle weight.  We have calculated $m^{\ast}/m$ as a function of $n$ at $U=7.0$.  The result is presented in Fig. \ref{fig-meffn}.  The mass enhancement factor for the $\beta$-Mn lattice shows a large peak at $n=1.1$.  The peak develops further, and becomes even more than ten when $U$ is increased from 7.0 to 7.5.  
These results are consistent with the large electronic specific heat found in $\beta$-Mn, because the electron number $n=1.1$ in the single band model corresponds to the $d$ electron number $n_{d}=5.5$ according to the five-fold equivalent band model for transition metals~\cite{kake86}.
The mass enhancement factor for the bcc structure is also enhanced around $n=1.0$.  

The enhancements of $m^{\ast}/m$ mentioned above are explained by the behavior of the particle and hole excitation functions, $\lambda_{1}^{(i)}(0)$ and $\lambda_{2}^{(i)}(0)$ in Eq. (\ref{qpz}):
\begin{equation}
Z = 1 -  
\sum_{i={\rm I},{\rm II}} \frac{ \rho_{i}(0) }{ \rho(0) } \,
\frac{ \lambda^{(i)}_{1}(0) + \lambda^{(i)}_{2}(0) }
{ 1 + C_{i} U_{i}^{2} \tilde{\eta}^{2}_{i} } \ ,
\label{qpz2}
\end{equation}
\begin{align} 
\lambda^{(i)}_{1}(0) =  U_{i}^{2} \tilde{\eta}^{2}_{i} 
\int^{0}_{-\infty} d\epsilon_{1} \int^{0}_{-\infty} d\epsilon_{3} \int_{0}^{\infty} d\epsilon_{4} \frac{ \rho_{i}(\epsilon_{1}) \, \rho_{i}(\epsilon_{3}) \, \rho_{i}(\epsilon_{4}) }
{(|\epsilon_{4}| + |\epsilon_{3}| + |\epsilon_{1}| + |\epsilon^{(i)}_{c}|)^{2} } \ ,
\label{lambda12}
\end{align}
\begin{align} 
\lambda^{(i)}_{2}(0) =  U_{i}^{2} \tilde{\eta}^{2}_{i} 
\int_{0}^{\infty} d\epsilon_{2} \int^{0}_{-\infty} d\epsilon_{3} \int_{0}^{\infty} d\epsilon_{4}  \frac{ \rho_{i}(\epsilon_{2}) \, \rho_{i}(\epsilon_{3}) \, \rho_{i}(\epsilon_{4}) }
{(|\epsilon_{4}| + |\epsilon_{3}| + |\epsilon_{2}| + |\epsilon^{(i)}_{c}|)^{2} } \ .
\label{lambda22}
\end{align}
Here we omitted the spin suffix $\sigma$ for brevity.

In the case of the bcc, the anomalous central peak in the bcc DOS is located just on the Fermi level at $n=1.0$.  Since the DOS $\rho_{i}(\epsilon)$ in $\lambda_{1}^{(i)}(0)$ and $\lambda_{2}^{(i)}(0)$ take large values near the low energy region, both $\lambda_{1}^{(i)}(0)$ and $\lambda_{2}^{(i)}(0)$ are enhanced at $n=1.0$, and thus $m^{\ast}/m$ shows a peak there.  We note that the peak is characteristic of the NN hopping model on the bcc lattice.

In the $\beta$-Mn lattice, the band widths of the LDOS are reduced by 30\% as compared with the fcc and the bcc cases.  The enhancement of the LDOS due to the band narrowing increases $\lambda_{1}^{(i)}(0)$ and $\lambda_{2}^{(i)}(0)$ according to the scaling relation $\lambda_{1}^{(i)}(0), \lambda_{2}^{(i)}(0) \varpropto W^{-2}$, where $W$ is a band width of the LDOS.  Moreover the Fermi level is located in the 3rd-peak region in the DOS when $n=1.1$ (see Fig. \ref{fig-dos0}) , and thus $\lambda_{1}^{(i)}(0)+\lambda_{2}^{(i)}(0)$ are further enhanced there.  It should be noted that the 3rd peak appears in the LDOS at both sites I and II, thus $m^{\ast}/m$ is enhanced by the electrons on both sites. 
The band narrowing effects and the 3rd peak on the Fermi level create the large peak of $m^{\ast}/m$ on the $\beta$-Mn lattice at $n=1.1$.

%
%
%---------------------------------------------------------------------
%---------------------------------------------------------------------
%
%

\section{Summary}

We have investigated the electronic structure and electron correlations of the single-band Hubbard model on the $\beta$-Mn type lattice on the basis of the momentum-dependent local ansatz (MLA) wavefunction method. 

The $\beta$-Mn lattice has 20 atoms in the unit cell with two inequivalent sites I and II.  Solving the eigen-value equations for the non-interacting system with the transfer integrals up to the 6 NN, we found that 
calculated band width of the $\beta$-Mn lattice is smaller than those of the fcc and the bcc with the NN transfer integrals by 30\%, so that the electrons on the $\beta$-Mn lattice are significantly correlated as compared with those on the other two lattices.  The density of states (DOS) for the $\beta$-Mn lattice consists of three main peaks: the highest peak near the top of the band, the 2nd highest peak at lower energy, and the 3rd peak near the Fermi level of the electron number $n=1.1$.
We also found that calculated local densities of states (LDOS) on site II are narrower than those on site I.  Thus, site II electrons are more localized, and the site II is ionized positively (negatively) for $n < 0.6$ ($n > 0.6$).

We calculated the correlation energy $\epsilon_{c}$ vs $n$ curves on the $\beta$-Mn, the fcc, and the bcc lattices for a typical value of Coulomb interaction energy of transition metals $U=7.0$.  We found that the correlation energy gain $|\epsilon_{c}|$ of the $\beta$-Mn lattice is the largest among the three structures, as expected from the narrow-band character of electrons on the $\beta$-Mn lattice.  Correlated electrons on the $\beta$-Mn lattice also cause a significant suppression of charge fluctuations $\langle (\delta n_{i})^{2} \rangle^{1/2}$, and a dip in the $\langle (\delta n_{i})^{2} \rangle^{1/2}$ vs $n$ curve at $n=1.1$.

We have investigated the momentum distribution function (MDF) on the $\beta$-Mn lattice.  At half-filling, the deviation from the Fermi distribution function rapidly increases with increasing Coulomb interaction $U$.  Accordingly, the quasiparticle weight obtained from the jump of the MDF at the Fermi level decreases with increasing $U$. 
We obtained the critical Coulomb interaction $U_{c2}=8.8$ for the $\beta$-Mn lattice, which is much smaller than the fcc one $15.0$.
 
Calculated mass enhancement factors $m^{\ast}/m$ of the $\beta$-Mn lattice are larger than those of the fcc irrespective of the average electron number $n$.  In particular, we found that $m^{\ast}/m$ of the $\beta$-Mn lattice shows a large peak at $n=1.1$ when the electron number $n$ is varied at $U=7.0$.  The peak is higher than the bcc one and even becomes more than ten when we slightly increase $U$.
We clarified that the large peak of the mass enhancement on the $\beta$-Mn lattice is caused by the band narrowing and the considerably large 3rd peak in the DOS at the Fermi level of $n=1.1$.  

The Mott-Hubbard type large mass enhancement on the $\beta$-Mn lattice found at $n=1.1$ in the present calculations may explains anomalous electronic specific heat coefficient of $\beta$-Mn at low temperatures, because the electron number $n=1.1$ corresponds to the $d$ electron number $n_{d} = 5.5$ according to the five-fold equivalent band model for transition metals~\cite{kake86}.  Correlated electron picture of $\beta$-Mn obtained in the present work is consistent with the Mott-Hubbard type single particle excitation spectra of $\gamma$-Mn obtained by the first principles dynamical CPA (coherent potential approximation) at high temperatures~\cite{kake09}, the XPS (X-ray photoelectron spectroscopy) data~\cite{bier01} for $\gamma$-Mn on the Cu${}_{3}$Au(100), and the BIS (bremsstrahlung isochromat spectroscopy) data~\cite{speier84} for $\alpha$-Mn at room temperature.

We adopted the single-band Hubbard model in the present work.  The quantitative aspects of such a simple model has not yet been clarified for the $\beta$-Mn type lattice.
In order to verify the present result showing a large electronic specific heat coefficient, we have to perform more realistic calculations using the first-principles MLA.~\cite{kake16,chan16,kake16-2,kake17,kake21}  The stability of the paramagnetic state has also to be examined.  The band calculations~\cite{hafner03} suggest that a ferrimagnetic state or more complex antiferromagnetic state is stable for $\beta$-Mn in contradiction to the experimental observation of the paramagnetism.  The Hund-rule coupling energy in the paramagnetic state, however, is missing in these calculations.  The Hund-rule energy may stabilize the paramagnetism, as has recently been found in the $\epsilon$-Fe system.~\cite{pour14}  It is highly desired to apply the first-principles MLA to the $\beta$-Mn system since the theory takes into account the Hund-rule correlations in the paramagnetic state.

%\begin{acknowledgment}
%\acknowledgment
%\end{acknowledgment}
\clearpage

\appendix

\section{Matrix Elements in the Correlation Energy}

 In this Appendix, we present the matrix elements $\langle \tilde{H} \tilde{O}_{i} \rangle_{0} = \langle \tilde{O}^{\dagger}_{i} \tilde{H} \rangle_{0}^{\ast}$, $\langle \tilde{O}^{\dagger}_{i} \tilde{H} \tilde{O}_{i} \rangle_{0}$, and 
$\langle \tilde{O}^{\dagger}_{i} \tilde{O}_{i} \rangle_{0}$ in the correlation energy (\ref{corr3}).
\begin{align}
\langle \tilde{H} \tilde{O}_{i} \rangle_{0} = U_{i} \sum_{\kappa_{1}\kappa_{2} \kappa^{\prime}_{1}\kappa^{\prime}_{2}} 
|\langle \kappa^{\prime}_{2} | i \rangle_{\downarrow}|^{2} 
|\langle \kappa_{2} | i \rangle_{\downarrow}|^{2} 
|\langle \kappa^{\prime}_{1} | i \rangle_{\uparrow}|^{2} 
|\langle \kappa_{1} | i \rangle_{\uparrow}|^{2} 
\tilde{f}_{\kappa^{\prime}_{2}\kappa_{2}\kappa^{\prime}_{1}\kappa_{1}} \eta^{(i)}_{\kappa^{\prime}_{2}\kappa_{2}\kappa^{\prime}_{1}\kappa_{1}} \ ,
\label{app-ho}
\end{align}
\begin{align}
\langle \tilde{O}^{\dagger}_{i} \tilde{H} \tilde{O}_{i} \rangle_{0} & = \sum_{\kappa_{1}\kappa_{2} \kappa^{\prime}_{1}\kappa^{\prime}_{2}} 
|\langle \kappa^{\prime}_{2} | i \rangle_{\downarrow}|^{2} 
|\langle \kappa_{2} | i \rangle_{\downarrow}|^{2} 
|\langle \kappa^{\prime}_{1} | i \rangle_{\uparrow}|^{2} 
|\langle \kappa_{1} | i \rangle_{\uparrow}|^{2} 
\tilde{f}_{\kappa^{\prime}_{2}\kappa_{2}\kappa^{\prime}_{1}\kappa_{1}} 
\eta^{(i)\ast}_{\kappa^{\prime}_{2}\kappa_{2}\kappa^{\prime}_{1}\kappa_{1}} 
\nonumber \\
& \hspace*{10mm} \times \Bigg[ \ 
\Delta E_{\kappa^{\prime}_{2}\kappa_{2}\kappa^{\prime}_{1}\kappa_{1}}
\eta^{(i)}_{\kappa^{\prime}_{2}\kappa_{2}\kappa^{\prime}_{1}\kappa_{1}} 
\nonumber \\
& \hspace*{17mm}  + U_{i} \,
\biggl\{ \,
\sum_{\kappa_{3}\kappa_{4}} |\langle \kappa_{3} | i \rangle_{\uparrow}|^{2} 
|\langle \kappa_{4} | i \rangle_{\downarrow}|^{2} f(\epsilon_{\kappa_{3\uparrow}})f(\epsilon_{\kappa_{4\downarrow}})
\eta^{(i)}_{\kappa^{\prime}_{2}\kappa_{4}\kappa^{\prime}_{1}\kappa_{3}}
\nonumber \\
& \hspace*{24mm} - 
\sum_{\kappa^{\prime}_{3}\kappa_{4}} 
|\langle \kappa^{\prime}_{3} | i \rangle_{\uparrow}|^{2} 
|\langle \kappa_{4} | i \rangle_{\downarrow}|^{2} f(-\epsilon_{\kappa^{\prime}_{3\uparrow}})f(\epsilon_{\kappa_{4\downarrow}})
\eta^{(i)}_{\kappa^{\prime}_{2}\kappa_{4}\kappa^{\prime}_{3}\kappa_{1}} 
\nonumber \\
& \hspace*{24mm} - 
\sum_{\kappa_{3}\kappa^{\prime}_{4}} 
|\langle \kappa_{3} | i \rangle_{\uparrow}|^{2} 
|\langle \kappa^{\prime}_{4} | i \rangle_{\downarrow}|^{2} 
f(\epsilon_{\kappa_{3\uparrow}})f(-\epsilon_{\kappa^{\prime}_{4\downarrow}})
\eta^{(i)}_{\kappa^{\prime}_{4}\kappa_{2}\kappa^{\prime}_{1}\kappa_{3}}
\nonumber \\
& \hspace*{24mm} + 
\sum_{\kappa^{\prime}_{3}\kappa^{\prime}_{4}} 
|\langle \kappa^{\prime}_{3} | i \rangle_{\uparrow}|^{2} 
|\langle \kappa^{\prime}_{4} | i \rangle_{\downarrow}|^{2} 
f(-\epsilon_{\kappa^{\prime}_{3\uparrow}})
f(-\epsilon_{\kappa^{\prime}_{4\downarrow}})
\eta^{(i)}_{\kappa^{\prime}_{4}\kappa_{2}\kappa^{\prime}_{3}\kappa_{1}}
\biggl\}  
\Bigg] \ ,
\label{app-oho}
\end{align}
\begin{align}
\langle \tilde{O}^{\dagger}_{i} \tilde{O}_{i} \rangle_{0} = \sum_{\kappa_{1}\kappa_{2} \kappa^{\prime}_{1}\kappa^{\prime}_{2}} 
|\langle \kappa^{\prime}_{2} | i \rangle_{\downarrow}|^{2} 
|\langle \kappa_{2} | i \rangle_{\downarrow}|^{2} 
|\langle \kappa^{\prime}_{1} | i \rangle_{\uparrow}|^{2} 
|\langle \kappa_{1} | i \rangle_{\uparrow}|^{2} 
\tilde{f}_{\kappa^{\prime}_{2}\kappa_{2}\kappa^{\prime}_{1}\kappa_{1}} |\eta^{(i)}_{\kappa^{\prime}_{2}\kappa_{2}\kappa^{\prime}_{1}\kappa_{1}}|^{2} \ .
\label{app-oo}
\end{align}
Here $\langle \kappa | i \rangle_{\sigma}$ denotes the overlap integral between the localized orbital $|i \rangle_{\sigma}$ and the Bloch state 
$|\kappa \rangle_{\sigma}$.
$\tilde{f}_{\kappa^{\prime}_{2}\kappa_{2}\kappa^{\prime}_{1}\kappa_{1}}=f(-\epsilon_{\kappa^{\prime}_{2\downarrow}})f(\epsilon_{\kappa_{2\downarrow}})f(-\epsilon_{\kappa^{\prime}_{1\uparrow}})f(\epsilon_{\kappa_{1\uparrow}})$, and $f(\epsilon)$ is the Fermi distrubution function at zero temperature.
Moreover, $\Delta E_{\kappa^{\prime}_{2} \kappa_{2} \kappa^{\prime}_{1} \kappa_{1}}$ is defined by
$\Delta E_{\kappa^{\prime}_{2} \kappa_{2} \kappa^{\prime}_{1} \kappa_{1}} 
= \epsilon_{\kappa^{\prime}_{2} \downarrow} - \epsilon_{\kappa_{2} \downarrow}
+ \epsilon_{\kappa^{\prime}_{1} \uparrow} - \epsilon_{\kappa_{1} \uparrow}$.

\end{document}